\documentclass[a4paper,11pt]{article}
\usepackage{jinstpub} 
\usepackage{lineno}

\usepackage{amsmath}
\usepackage{siunitx}
\usepackage{subcaption}
\usepackage[acronym]{glossaries}
\usepackage{acronym}
\usepackage{booktabs}
\usepackage{multirow}
\usepackage{makecell}
\usepackage{setspace}

\title{Using a radiopharmaceutical production cyclotron for radiobiological research}

\author[a,1]{Eva Kasanda,\note{Corresponding author.}}
\author[b]{Jan Gruber,}
\author[a,c,d]{Pierluigi Casolaro,}
\author[a]{Lars Eggimann,}
\author[a]{Alexander Gottstein,}
\author[a]{Isidre Mateu,}
\author[b]{Paolo Pellicioli,}
\author[e]{Maria Vittoria Rossi,}
\author[a,d]{Paola Scampoli,}
\author[b]{Cristian Fernandez-Palomo}
\author[a]{and Saverio Braccini}

\affiliation[a]{Albert Einstein Center for Fundamental Physics (AEC), Laboratory for High Energy Physics (LHEP), University of Bern,
Sidlerstrasse 5, 3012 Bern, Switzerland}
\affiliation[b]{Institute of Anatomy, University of Bern, 
Baltzerstrasse 2, 3012 Bern, Switzerland}
\affiliation[c]{INFN Sezione di Napoli,
Complesso Univ. Monte S. Angelo, 80126 Napoli, Italy}
\affiliation[d]{Department of Physics ``Ettore Pancini'', University of Napoli Federico II,
Complesso Univ. Monte S. Angelo, 80126 Napoli, Italy}
\affiliation[e]{Department of Science and High Technology, Universit\`a degli Studi dell'Insubria,
Via Valleggio 11, 22100 Como, Italy}

\emailAdd{eva.kasanda@unibe.ch}

\abstract{The \ac{LHEP} and the Institute of Anatomy at the University of Bern are establishing a pre-clinical proton therapy research facility at the Bern University Hospital (Inselspital). The facility centers on the \ac{BMC}, an 18~MeV compact cyclotron designed for $\mu$A-range beam currents for radioisotope production. The \ac{BMC} can be adapted as a highly accessible and customizable platform for pre-clinical proton therapy research, enabling advanced modalities such as ultra-high dose rate (FLASH) radiotherapy and \ac{SFRT}. Human keratinocytes (HaCaT) and mouse melanoma (B16-F10) cell lines were irradiated with 8.14(29)MeV protons at conventional dose rates and with 225kV X-rays at doses from 0 to 8~Gy. Clonogenic survival assays were analyzed using the \ac{LQ} model, and \ac{RBE} values were derived by comparing the proton and X-ray doses required for 0.1 survival. Proton irradiation yielded steeper survival curves than X-rays, indicating higher cytotoxicity. B16-F10 cells were most sensitive, with an \ac{RBE} of 1.34 at 0.1 survival, while HaCaT cells showed an \ac{RBE} of 1.21. These results confirm enhanced proton effectiveness in cell killing and demonstrate that the \ac{BMC} provides a reliable platform for radiobiological studies. Establishing in vitro proton irradiation capability marks a key milestone toward developing a comprehensive pre-clinical radiobiology research facility in Bern. Ongoing upgrades will further refine dosimetry and proton delivery, paving the way for future FLASH and proton \ac{SFRT} studies.}


\begin{document}
\maketitle
\flushbottom

\section{Introduction}

\ac{PT} is a form of \ac{RT} that uses protons, as opposed to the X-rays used in conventional \ac{RT}, to deliver a localized dose of ionizing radiation to a tumour. \ac{PT} has been in clinical use and steadily gaining popularity for decades due to its superior normal tissue sparing capabilities relative to conventional \ac{RT}.

Recent advancements have brought proton therapy to the threshold of a new era, with the first clinical applications of proton FLASH demonstrating the feasibility of delivering ultra-high dose rate proton treatments \cite{Matuszak2022, Mascia2023}. These pioneering efforts highlight the strong potential of FLASH to widen the therapeutic window, but they also underline a key limitation: the scarcity of radiobiological data that can clarify the mechanisms underlying normal tissue sparing and help establish optimal treatment parameters. In parallel, Spatially Fractionated Radiation Therapy (\ac{SFRT}) has also been demonstrated to significantly increase the tolerance of normal tissues while enhancing tumour control \cite{Yan2020, Mali2024}. However, as with FLASH, more pre-clinical radiobiological studies are needed to fully understand the biological mechanisms at play and to establish robust clinical protocols. 

Researchers investigating these techniques would benefit from an accessible proton facility able to offer direct comparisons between different treatment configurations. As a compact medical cyclotron designed for use at high currents for commercial radioisotope production, the Bern Medical Cyclotron (\ac{BMC}) offers such potential. With its extended \ac{BTL} accessible from a separate research bunker, the \ac{BMC} is in a position to be adapted for use as a highly customizable and easily accessible proton therapy research facility with FLASH capabilities, led by the Laboratory for High Energy Physics (\ac{LHEP}) at the University of Bern. In collaboration with the University of Bern’s Institute of Anatomy, which provides access to conventional and high-dose-rate X-ray sources and a radiobiology laboratory, the overarching goal is to prepare the \ac{BMC} laboratory towards the final goal of pre-clinical studies at FLASH dose rates and with proton \ac{SFRT}. The 18~MeV extracted beam energy enables in-vitro cell irradiations, as well as in-vivo investigations of small superficial tumours. This will enable direct radiobiological comparison studies between protons and photons across a wide range of dose rates.

To investigate the feasibility of performing pre-clinical proton therapy studies at the \ac{BMC}, a preliminary beam shaping setup to control the extracted dose rate and deliver a field large enough to uniformly irradiate a cell flask was developed. The preliminary beam shaping configuration and dose monitoring procedure are outlined in Kasanda et. al \cite{Kasanda2026-submitted}. 

In this work, we present initial radiobiology results from the \ac{BMC}, comparing the survival curves of two cell lines irradiated with protons and photons. For this proof-of-concept study, all irradiations were performed at conventional dose rates. 

\section{Materials and Methods}

\subsection{Proton Irradiation}
\subsubsection{Experimental Facility}
The \ac{BMC} is an IBA Cyclone 18/18 HC cyclotron housed at the Bern University Hospital (Inselspital), in Switzerland \citep{Braccini2013}. The cyclotron facility was developed as a result of a collaboration between the University of Bern, Inselspital, and private investors. This collaboration allows joint use of the accelerator for research in medical applications of particle physics alongside routine production of radiopharmaceuticals for positron emission tomography (PET). 

The \ac{BMC} is equipped with a \ac{BTL}: a \SI{6.5}{\meter}-long beamline with steering and focusing magnets. The \ac{BTL} ends in a dedicated research bunker (called the \ac{BTL} bunker), which can be accessed without exposure to activity from the cyclotron and radiopharmaceutical production targets, which are all housed within the adjacent cyclotron bunker \citep{Braccini2013}. 

\begin{figure}[h]
    \centering
    \includegraphics[width=\textwidth]{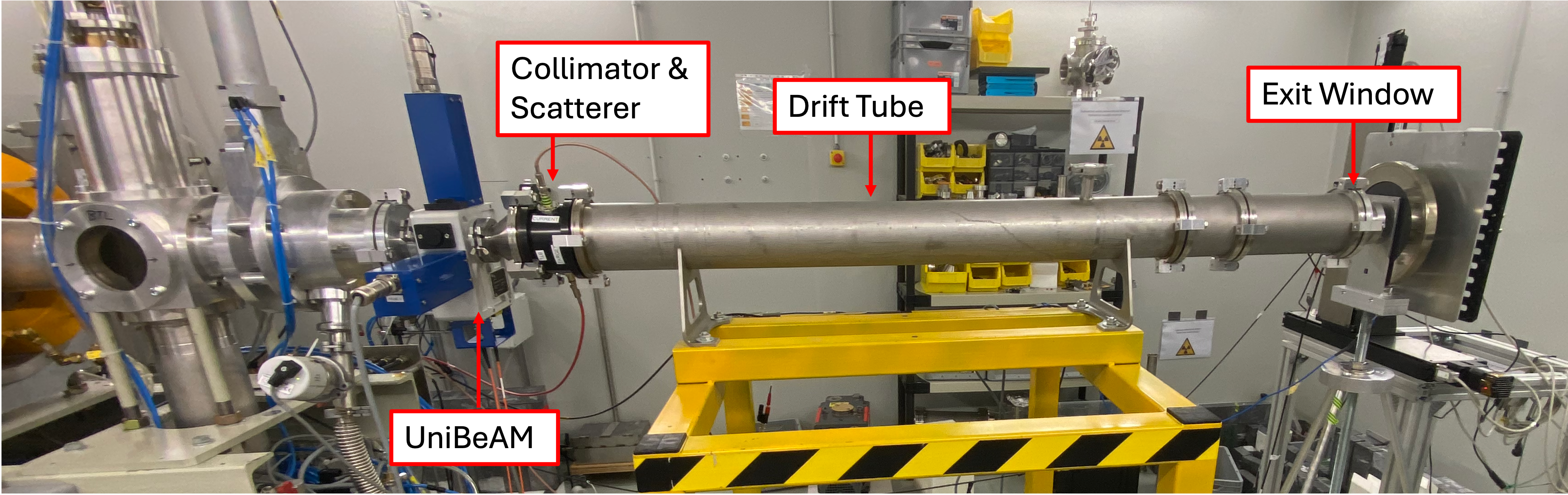}
    \caption{\ac{BTL} of \ac{BMC} in configuration for proton therapy studies, as described in Kasanda et al \cite{Kasanda2026-submitted}, including the UniBeAM beam profiler \citep{Potkins2017} and beam shaping components. }
    \label{fig:btl}
\end{figure}

The \ac{BMC} is able to deliver 18~MeV proton beams with currents up to \SI{150}{\micro A} ($>$~\SI{1000}{Gy/s}), making it well-suited for FLASH studies \citep{Auger2015}. Collimating and scattering the beam as described in Kasanda et al \cite{Kasanda2026-submitted} and shown in Figure \ref{fig:btl} allows currents as low as a few pA ($<$~\SI{0.05}{Gy/s}) to be delivered to a target in air, enabling direct comparisons between conventional and FLASH \ac{PT}. While the relatively low energy and penetration depth of the extracted protons impose some constraints on the radiobiological studies that can be performed, they also significantly simplify beam collimation for \ac{SFRT} investigations.

\subsection{Target Mounting and Irradiation Procedure}

\begin{figure}[h]
    \centering
    \begin{subfigure}[b]{0.45\textwidth}
        \includegraphics[width=\textwidth]{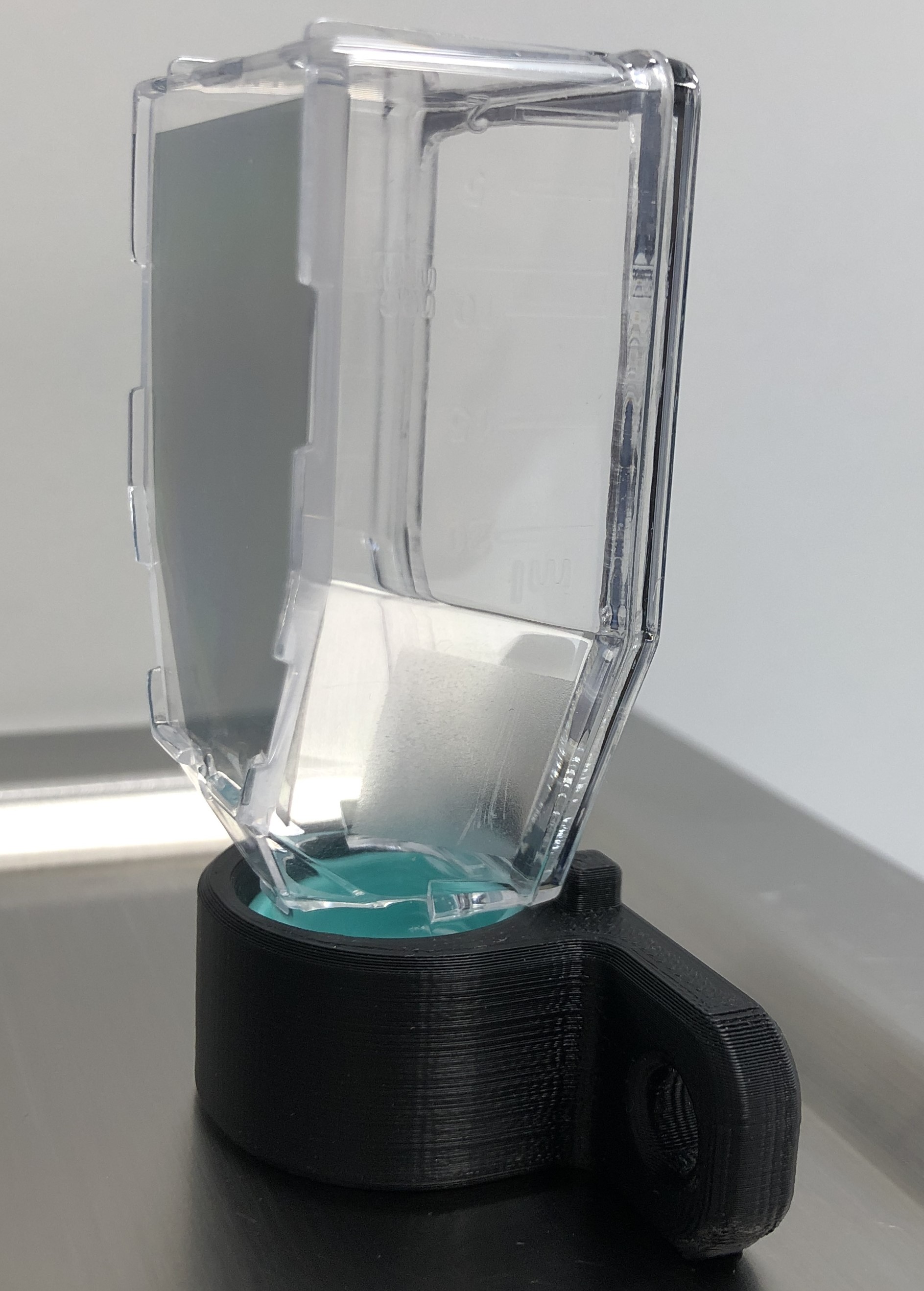}
        \caption{~}
        \label{fig:subfig:flask}
    \end{subfigure}%
    \hfill
    \begin{subfigure}[b]{0.47\textwidth}
        \includegraphics[width=\textwidth]{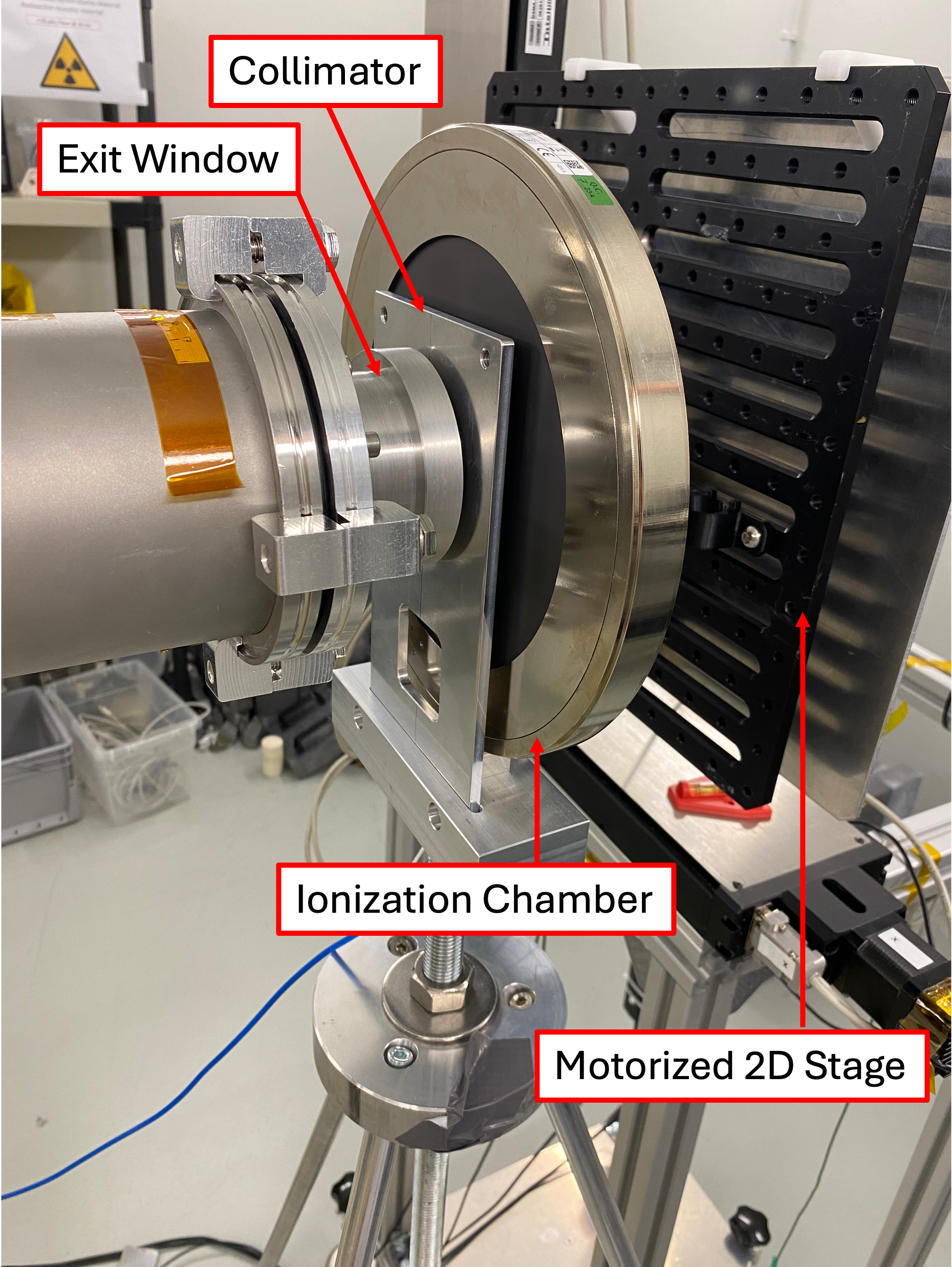}
        \caption{~}
        \label{fig:subfig:target}
    \end{subfigure}%
    \caption{Experimental setup for in-vitro irradiations. (a) Cell culture flask used for in-vitro irradiations and 3D-printed holder. A calibration EBT-3 film is visible positioned on the inner surface of the flask on which the cells are attached. (b) Target mounting area, including ionization chamber and 2D stage for target mounting. The collimator shown in this picture is a $1\times1$~cm$^2$ aperture that was used for calibration purposes only and was removed to irradiate the cells.}
    \label{fig:target-invitro}
\end{figure}

For in-vitro studies, the lucite cell flask depicted in Figure \ref{fig:subfig:flask} was used. The face on which cell colonies are grown is \SI{1.5}{\milli\meter} thick. A 3D-printed holder was used to secure the flask in a vertical orientation on a remotely controlled 2D stage with sub-millimeter positioning accuracy. Taking into consideration the in-beam ionization chamber positioned in air between the extraction window and the target (shown in Figure \ref{fig:subfig:target}), the beam energy extracted into air and at the entrance of the cell layer in this configuration have been measured to be \SI{15.54(12)}{MeV} and \SI{8.14(29)}{MeV}, respectively \citep{Gottstein2025}. This result is in agreement with LISE++ simulations \citep{Tarasov2008} of the beamline, which predicted a beam energy of \SI{7.88(29)}{MeV} at the entrance of the cell layer in this configuration. The experimentally measured entrance energy corresponds to a dose-weighted \ac{LET} of \SI{5.425(25)}{keV/\micro\meter} inside the cells. The change in Linear Energy Transfer (\ac{LET}) throughout the cell layer is less than 0.5\%. Note that the dose-weighted LET is calculated from the distribution of proton energies impinging on the cell layer. Since the quoted uncertainty is dominated by energy straggling, this LET estimation does not include any uncertainties. 

Irradiation was performed approximately six hours after cell seeding. For transportation to the irradiation centers, the cells were placed in a polystyrene container equipped with heat packs. For proton irradiation, the long exposure times required the use of an incubator at the \ac{BMC} facility to maintain temperature control. During proton irradiation, the flasks were placed vertically with their lids facing down, as shown in Figure \ref{fig:experiment-sketch}. This vertical arrangement caused the culture medium to flow into the lids of the flasks, temporarily depriving the cells of direct contact with the medium. The impact of this configuration was quantified using a set of 6 control flasks held in the irradiation position for the duration of the longest irradiation. Due to the proton field being only \SI{55}{\milli\meter} in diameter and the cell flasks being approximately \SI{35}{\milli\meter} $\times$ \SI{45}{\milli\meter} in area, flasks were irradiated individually to ensure uniform dose delivery to the area of interest. 
Over the course of all proton irradiations, the average delivered dose rate was approximately \SI{1.9(8)}{Gy/min}. 

\subsection{Dosimetry}
After the proton beam is extracted into air, it passes through a PTW type 786 monitoring ionization chamber (IC) \citep{PTW}, which provides real-time feedback of the total proton current incident on the IC during irradiation. This IC has been calibrated to dose rate delivered to the target using EBT-3 Gafchromic films \citep{Ashland2023}. Since the lateral spread and energy loss of the extracted protons in air is significant in this region, the calibration must be repeated for each setup, as small changes in positioning can affect the relation between the IC signal response and the delivered dose. 

The dose response of the Gafchromic films has been calibrated using X-rays at multiple independent facilities. However, there are significant quenching effects impacting the response of the films to low-energy protons relative to X-rays which must be considered. 

The underresponse of EBT-3 films to protons in the Bragg Peak due to high \ac{LET} is well documented, and the relative efficiency of the films at the measured energy accounts for an underestimation of the dose by a factor of 0.908(22) \citep{SanchezParcerisa2021}. 

The cell layer is only approximately \SI{15}{\micro\meter} in thickness, while the polyimide laminate layer of the EBT-3 film is \SI{125}{\micro\meter} in thickness, resulting in additional energy loss before the particle beam reaches the active layer of the film. For high energy protons and X-rays, the dose gradient within the film is low enough that the dose at the entrance of the active layer is not very different from the dose at the entrance of the plastic lamination layer. However, for protons in the Bragg peak region, there is significant energy loss in the film before the active layer is reached, resulting in a higher dose impinging on the active layer than would be delivered to a thin layer of cells. At proton energies below 20 MeV, this caused a non-negligible increase in \ac{LET} through the lamination layer, resulting in an overestimation of the dose delivered to the cells. 

This effect has been estimated to account for an overestimation of the dose by a factor of 1.191(73), based on experimental measurements of the change in the dose response of EBT-3 films through additional layers of the \SI{125}{\micro\meter}-thick polyimide material that coats the active layer, as described in Kasanda et al \cite{Kasanda2026-submitted}.

\subsection{X-ray irradiation}

\subsubsection{X-ray Radiation source}
The X-RAD 225 cabinet (Precision X-ray Inc., Madison, USA) at the \ac{DBMR} of the University of Bern was used for X-ray exposure. X-ray irradiations were performed with maximum tube voltage of \SI{225}{\kilo V} and maximum current of  \SI{17.5}{\milli A}.
The X-ray beam was filtered using a \SI{0.3}{\milli\meter} copper filter and the cells were placed horizontally at a source-to-sample distance of \SI{30}{\centi\meter}. Irradiations were conducted at a dose rate of \SI{4.71}{Gy/min}.

\subsection{Irradiation Procedure}
\begin{figure}[h]
    \includegraphics[width=\textwidth]{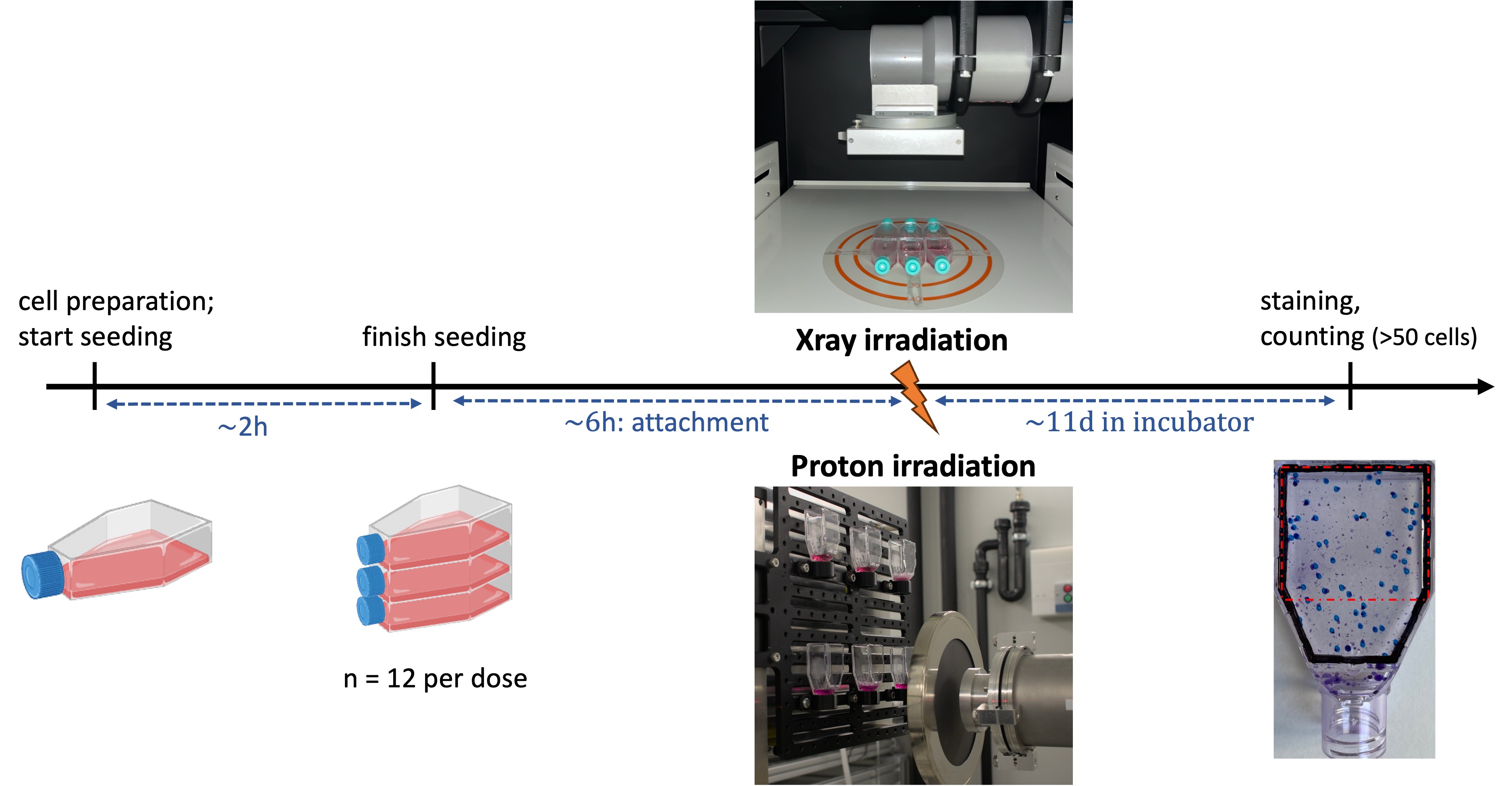}
    \caption{A schematic representation of the experimental setup, including the timeline and flask positioning for the irradiation procedure. During X-ray irradiation, the flasks were placed horizontally to ensure continuous contact between the cells and the culture medium. Six flasks were irradiated under these conditions. In contrast, proton irradiation required the flasks to be positioned vertically with the lids facing downward, so the cells were not in contact with the culture medium during the exposure period. Figure modified with permission from J. Gruber \cite{Gruber2026}.}
    \label{fig:experiment-sketch}
\end{figure}

The shorter exposure times for X-rays allowed the polystyrene container alone to maintain sufficient temperature control. The larger field size compared with the proton beam allowed the irradiation of multiple flasks at once, as depicted in Figure \ref{fig:experiment-sketch}. Additionally, the flasks could be placed horizontally such that the cells remained in contact with the culture medium throughout the irradiation.
For higher doses, irradiation was performed in sets of three flasks at a time. The intended dose ranged from 0 to 8 Gy, with the number of cells adjusted to each dose point. All irradiations were performed under controlled room temperature conditions at 22~$^\circ$C.

\subsection{Cell Lines}
\subsubsection{Cell line – HaCaT (human keratinocytes)}
The HaCaT cell line is composed of spontaneously immortalized ($<$ 140 passages) aneuploid human keratinocytes derived from histologically normal skin of a 62 year old Caucasian male with melanoma. Boukamp et al \cite{Boukamp1988} established this cell line and found that despite the altered and unlimited growth potential, the cells maintain normal epidermal differentiation, remaining nontumorigenic, and retain the capacity to reconstitute a well-structured epidermis after in vivo transplantation. The HaCaTs were purchased from Cell Lines Service GmbH, Eppelheim, Germany. They were regularly maintained with Dulbecco’s Modified Eagle Medium (DMEM; Gibco, catalog number: 11965-092) + 4.5g/L glucose + L-glutamine + phenol red + HEPES (Gibco, catalog number: 15630-056) supplemented with 10\% fetal bovine serum (FBS; Gibco, catalog number: 16000-044), without antibiotics.

\subsubsection{Cell line – B16-F10 (mouse melanoma)}
 The B16-F10 melanoma cell line is a subclone of the B16 mouse tumour line and is derived from lung metastases after intravenous injection of B16-F0 cells into a C57BL/6Cr mouse. After ten consecutive passages, the B16-F10 cell line was selected \citep{Nakamura2002}. They have a spindle-shaped, epithelial-like morphology and adherent growth. Different doubling times ranging from 14.2 to 20.1 hours have been reported \citep{Ohira1994}, \citep{Yerlikaya2008}. The B16-F10 cells are characterized by excessive melanin release and the potential to metastasize to the lung parenchyma in vivo \citep{Nakamura2002}. 
The B16-F10 cell line was purchased from LGC Standards GmbH, Germany, which acts as a distributor of ATCC, USA (Catalog number: CRL 6475). This cell line was cultivated with DMEM (Gibco, catalog number: 11995-056) + 4.5g/L glucose + L-glutamine + phenol red + sodium pyruvate + HEPES (Gibco) supplemented with 10\% FBS (Gibco), without antibiotics.

\subsection{Cell Culture Technique}
To prepare the experiments, the cell lines were seeded in multiple \SI{75}{\square\centi\meter} flasks with 25 mL of culture medium. Flasks were used at 90\% confluency and at cell passage numbers between 4 and 9, the cells were carefully rinsed with 20 mL Dulbecco's phosphate-buffered saline (DPBS; Gibco, catalog number: 14190-094). The cells were then detached with 10 mL of 0.05\% trypsin-EDTA solution (0.5\%) (Gibco, 10-fold concentration, catalog number: 15400-054). The cell suspension was then centrifuged at 260 g for 5 min. The supernatant was discarded, the cell pellet was resuspended with culture medium, and the cells were counted manually using a hemocytometer. Small \SI{12.5}{\square\centi\meter} flasks (BancLabo) were seeded in six replicates with the appropriate cell numbers for doses between 0 and 8 Gy and incubated for 5 to 6 hours in a humidified atmosphere with 5\% CO$_2$ at 37 $^\circ$C before irradiation. The cell numbers were established through preliminary X-ray experiments for each cell line and its corresponding dose points. These numbers were subsequently used in proton trials and adjusted based on deviations from the optimal fit to the linear-quadratic model.

\subsection{Clonogenic assay}
After irradiation, the flasks were immediately returned to the incubator at the Institute of Anatomy and were maintained in a humidified atmosphere of 5\% CO$_2$ at 37 $^\circ$C for a period sufficient to allow colony formation: 9 days for B16-F10 cells and 14 days for the HaCaT cell line. After the incubation period, colonies were stained with crystal violet and those with more than 50 cells were counted as survivors. 

\section{Results}
\subsection{Proton dose calculations}
The IC calibration yielded a dose-to-charge ratio of \SI{3.98}{MGy/C} for a field size of \SI{55}{\milli\meter} in diameter. This conversion factor was obtained in the exact configuration in which the cells were irradiated, and therefore encompasses geometric effects such as scattering and reductions in fluence induced by beamline elements. Additionally, since the EBT-3 films were mounted inside the flask at the position of the cell layer, the energy loss in the wall of the flask is also taken into consideration. 

\begin{figure}[h]
     \centering
     \begin{subfigure}[b]{0.59\textwidth}
         \includegraphics[width=\textwidth]{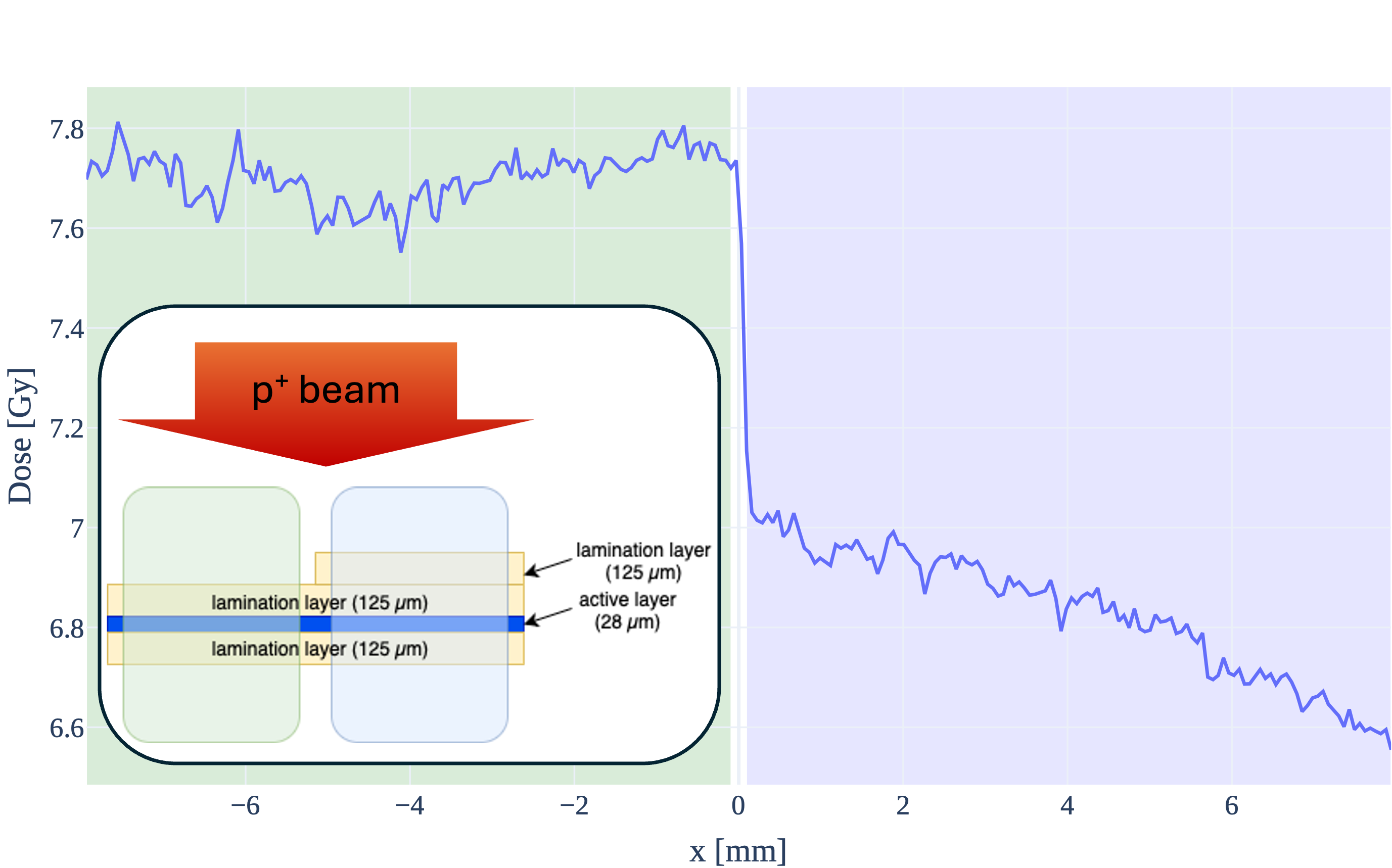}
         \caption{~}
         \label{fig:subfig:lamination}
     \end{subfigure}%
     \hfill
     \begin{subfigure}[b]{0.38\textwidth}
         \includegraphics[width=\textwidth]{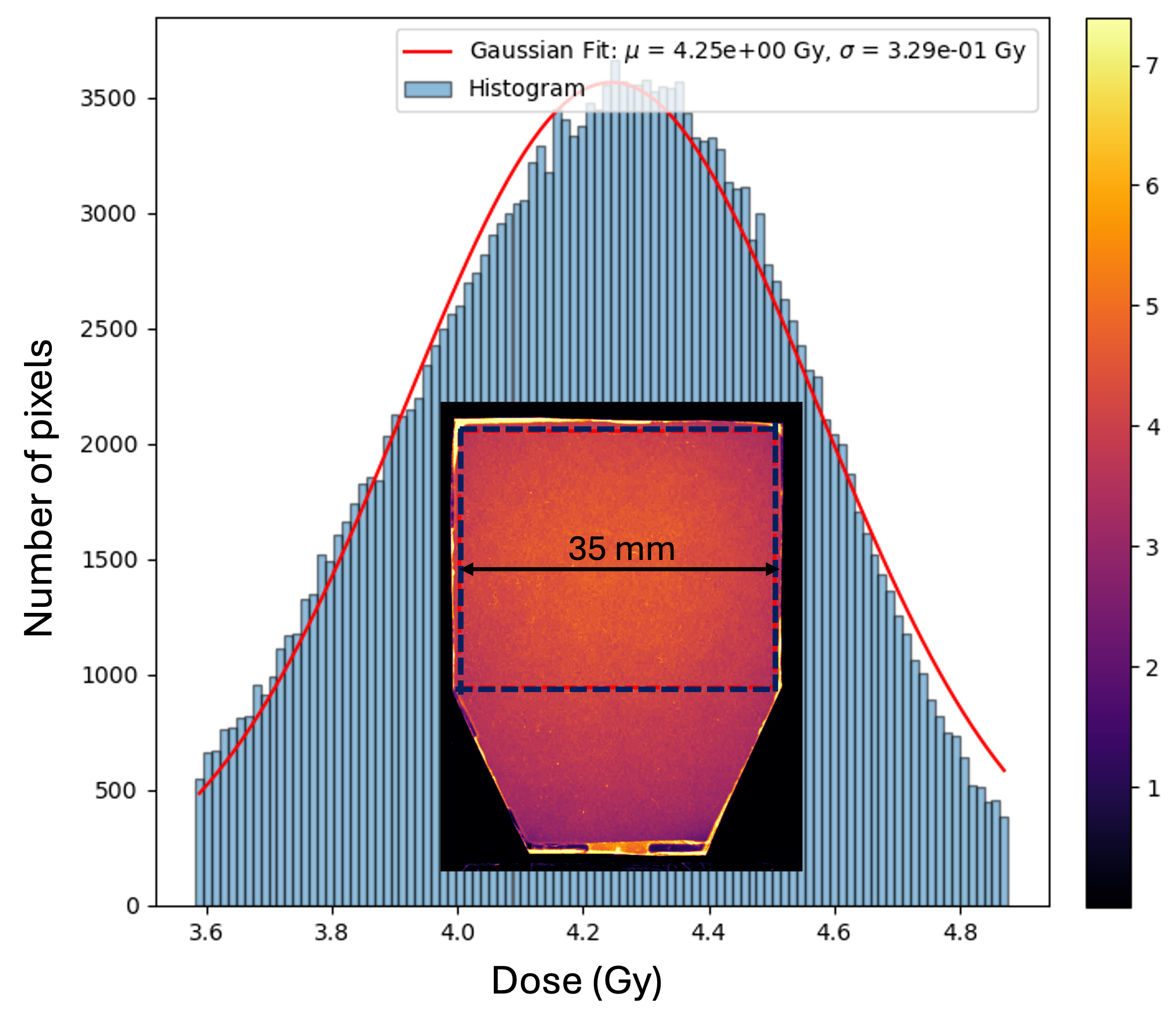}
         \caption{~}
         \label{fig:subfig:dosehist}
     \end{subfigure}%
     \caption{Experimental determination of correction factors and uncertainties in dose calculation. (a) Dose profile extracted from EBT-3 film with additional lamination layer in green shaded area, as depicted in sketch and described in Kasanda et al \cite{Kasanda2026-submitted}. The extracted dose ratio was 1.191(73). (b) Histogram of dose delivered to the region of the cell flask in which the colonies are counted (area outlined in dotted blue line in inset image). A Gaussian fit of the distribution yields a standard deviation of 7.7\%. }
     \label{fig:correctionfactors}
 \end{figure}

To quantify the dose overestimation due to proton energy loss in the Gafchromic film lamination layer, a measurement was performed in which an EBT-3 film was partially covered by a \SI{125}{\micro\meter} polyimide layer as depicted in Figure \ref{fig:subfig:lamination}, and irradiated with a flat beam at the position of the cell layer. Since the two portions of the film were irradiated with the same proton flux, the discrepancy in the dose read from each film reflects the difference in proton \ac{LET} before and after the polyimide layer of the film. As shown in the scanned image of the film depicted in Figure \ref{fig:subfig:lamination}, this effect is notable, and was calculated to account for an overestimation of the dose by a factor of 1.191(73). 

The passive scattering configuration described in \citep{Kasanda2026-submitted} enables a dramatic reduction in dose rate and increased field size relative to what can be achieved using focusing quadrupoles. Spatial constraints in the \ac{BTL} bunker limit the extracted beam profile to a dose uniformity of 7.7\% in the area of interest within the cell flask, as shown in Figure \ref{fig:subfig:dosehist}. 

 The overall proton dose uncertainty was determined to be 12.8\%, accounting for contributions from non-uniformities in the beam profile, film correction factors, and IC calibration uncertainties.

\subsection{Cell Irradiation results}
\begin{figure}[h!]
    \centering
     \includegraphics[width=0.8\textwidth]{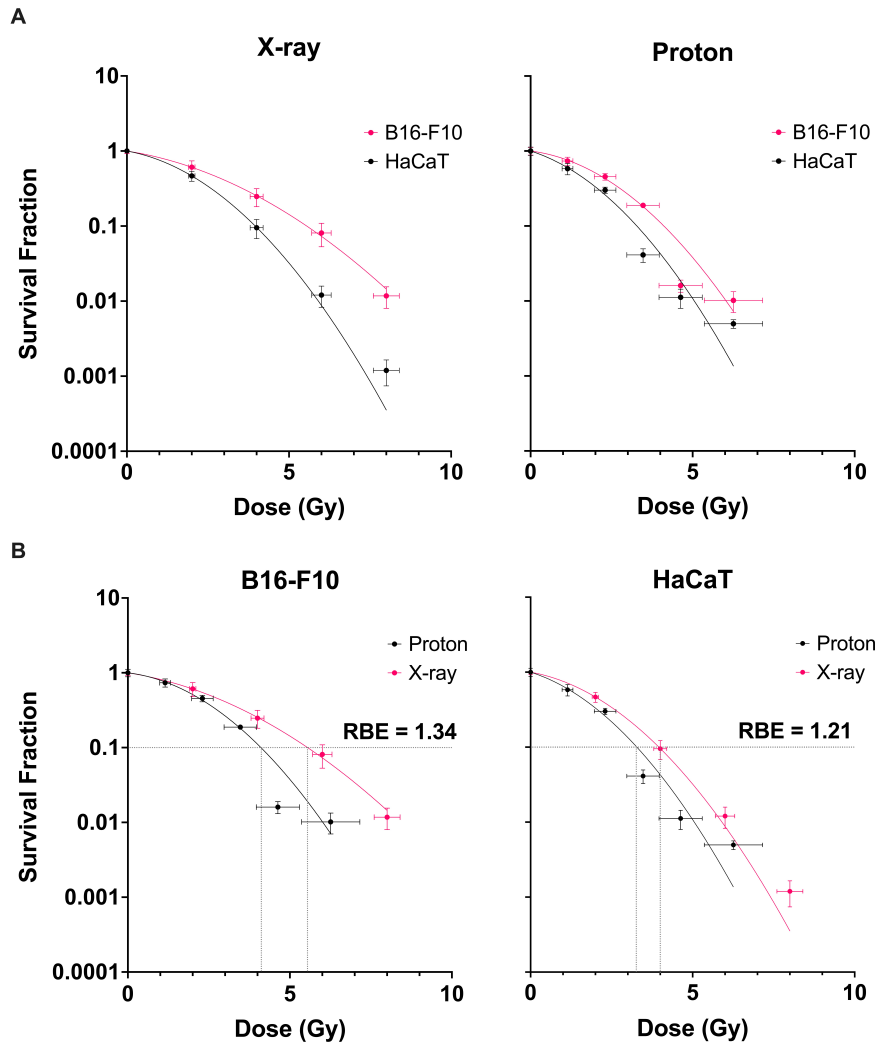}
    \caption{Comparison between the different cell lines after proton and X-ray irradiation (A) and dose dependent surviving fractions of the B16-F10 and HaCaT cells after irradiation with X-rays and protons (B). The survival curves show the best fit of the survival data to the \ac{LQ} model. The \ac{RBE} was calculated at 0.1 survival. Clonogenic survival was obtained from two independent experiments with 12 flasks in total. The data points are represented as the SEM and the horizontal error bars highlight the dose uncertainty. We assumed a 5\% uncertainty in the X-ray dose delivery. The overall proton dose uncertainty was 12.8\% (see Section \ref{sec:discussion}). Figure modified with permission from J. Gruber \cite{Gruber2026}.}
    \label{fig:survival_curves}
\end{figure}

Figure \ref{fig:survival_curves} panel A illustrates the survival curves of each cell line under X-ray and proton irradiation, highlighting distinct differences in radiosensitivity and survival between the two modalities. Panel A shows that HaCaT cells exhibit a steeper decline in survival compared to the B16-F10 cell line after protons and X-rays. This indicates that HaCaT cells are significantly more radiosensitive than the melanoma cell line, irrespective of the radiation type. 

Panel B shows the response of B16-F10 and HaCaT cells after being irradiated with X-rays or protons. The B16-F10 cell line exhibits a pronounced decrease in survival after proton irradiation compared to X-ray irradiation, highlighting the enhanced radiosensitizing effect of protons. This was quantified by a \ac{RBE} of 1.34 at 0.1 survival, indicating that proton radiation is 34\% more effective than X-rays and requires a lower dose to achieve the same reduction in survival.
HaCaT cells demonstrated a more pronounced decline in survival with increasing proton doses, as reflected by an \ac{RBE} of 1.21. This indicates a moderate radiosensitizing advantage of protons over X-rays in these cells. 

The best-fit parameter values of the \ac{LQ} model are presented in Table~\ref{tab:lq_excel_exact}. For B16-F10 cells, the $\alpha / \beta$ ratio was found to be 1.09 for protons and 3.2 for X-rays. For HaCaT cells, the ratio was higher for protons (2.58) than for X-rays (1.8).

\begin{table}[ht]
\centering
\caption{Best-fit values for the Linear-Quadratic model on HaCaT and B16-F10 cell line after X-ray and proton irradiation}
\label{tab:lq_excel_exact}
\renewcommand{\arraystretch}{1.2}
\setlength{\tabcolsep}{5pt}

\resizebox{\textwidth}{!}{%
\begin{tabular}{lcccccc}
\toprule
\multicolumn{1}{c}{\text{Cell Lines}} &
  \multicolumn{2}{c}{\text{X-ray LQ model}} &
  \text{Confidence limits (95\%)} &
  \multicolumn{2}{c}{\text{Proton LQ model}} &
  \text{Confidence limits (95\%)} \\
\midrule

\multirow{4}{*}{\textbf{HaCaT}} & $\alpha/\beta$ & 1.8&  & $\alpha/\beta$ & 2.58&  \\
 & $\alpha$ & 0.18& 0.08--0.26 & $\alpha$ & 0.31 & 0.21--0.41 \\
 & $\beta$ & 0.1 & 0.07--0.14 & $\beta$ & 0.12 & 0.07--0.17 \\
 & & & & & & \\

\multirow{3}{*}{\textbf{B16-F10}} & $\alpha/\beta$ & 3.2&  & $\alpha/\beta$ & 1.09&  \\
 & $\alpha$ & 0.16 & 0.09--0.22 & $\alpha$ & 0.12 & 0.06--0.18 \\
 & $\beta$ & 0.05 & 0.03--0.07 & $\beta$ & 0.11 & 0.08--0.13 \\
\bottomrule
\end{tabular}%
}
\end{table}

\section{Discussion}\label{sec:discussion}
This study aimed to evaluate the feasibility of conducting biological experiments at the \ac{BMC} facility at Inselspital, in Bern, and lay the groundwork for establishing a preclinical proton radiotherapy research platform supporting advanced applications, such as FLASH radiotherapy and \ac{SFRT}. As an initial step, in vitro experiments were performed using conventional dose rates and a uniform proton beam to validate the experimental setup. Clonogenic survival assays were conducted with HaCaT (human keratinocytes) and B16-F10 (mouse melanoma) cell lines. The survival fractions were normalized to the untreated controls and the dose-response relationships were analyzed using the \ac{LQ} model to generate cell-specific survival curves.

\subsection{Dose uncertainties}

For X-ray irradiations with the X-RAD 225 cabinet, a standard dose uncertainty of 5\% was assumed for survival curve fitting. The overall uncertainty in the delivered proton dose is 12.8\%, including contributions from: 
\begin{itemize}
    \item non-uniformities in the proton fluence delivered to the region of the cell flask in which the surviving cell colonies are counted (7.7\%)
    \item uncertainties in the interpretation of the dose delivered to the calibration film due to the high \ac{LET} of the protons (10.0\%)
    \item the uncertainty in the fitting parameters for the calibration of the ionization chamber (1.3\%)
\end{itemize}

 The uncertainty in the ionization chamber charge collection is considered negligible relative to the uncertainties listed above.

The dose response of the Gafchromic EBT-3 films used in the current setup has been calibrated using X-rays at multiple independent facilities . However, there are significant quenching effects impacting the response of the films to low-energy protons relative to X-rays that must be considered \cite{Kasanda2026-submitted}. The correction factors that are needed to compensate for these effects vary with proton energy, creating significant uncertainties in dose calculations. 

\subsection{Other facilities with similar capabilities}
To our knowledge, the BMC is one of 6 facilities worldwide that have documented setups for pre-clinical external beam therapy studies with protons using a radiopharmaceutical cyclotron. The key features of these setups are summarized in Table \ref{tab:cyclotrons}.

\begin{table}[htbp]
\centering

\caption{Medical/radiopharmaceutical cyclotrons adapted for external beam radiobiology studies.}
\label{tab:cyclotrons}
\begin{scriptsize}

\begin{tabular}{p{1.6cm}p{1.5cm}p{1.4cm}p{1.5cm}p{2.5cm}p{3.0cm}}

\toprule
\textbf{Facility} &
\textbf{Country} &
\makecell{\textbf{Energy}\\\textbf{at target}} &
\makecell{\textbf{Reported}\\\textbf{dose rate}\\\textbf{(Gy/s)}} &
\makecell{\textbf{Field diameter}\\\textbf{(dose error)}} &
\textbf{Key references}\\
\midrule

CYRCé &
France &
Few keV -- 24 MeV &
$10^{-5}$ -- 100 &
24 mm ($\pm2\%$) &
\cite{Fabbrizi2025}
\\

ARRONAX &
France &
61 MeV &
0.001 -- 40 &
15 mm ($\pm5\%$) &
\cite{Evin2024, Ralite2026}
\\

KIRAMS &
S. Korea &
34.9 MeV &
$\sim$0.5 &
$>$50 mm &
\cite{jung2012}
\\

CNA &
Spain &
12.8 MeV &
0.03 -- 0.05 &
35 mm ($\pm10\%$) &
\cite{BarattoRoldn2020}
\\

ICNAS &
Portugal &
14 MeV &
10 -- 16 &
21 mm &
\cite{Teixeira2026}
\\

BMC &
Switzerland &
15.54 MeV &
0.01 -- 1000+ &
55 mm ($\pm7.7\%$) &
\cite{Kasanda2026-submitted}
\\

\bottomrule
\end{tabular}

\end{scriptsize}
\end{table}

While CYRCé and ARRONAX have dedicated radiobiology beamlines, CNA and ICNAS are the facilities comparable to our own. Like the BMC, both are adding new functionality to an existing IBA 18 MeV Medical Cyclotron, which is currently being used for radiopharmaceutical production. Each facility has its own approach to beam shaping and monitoring. We are the first of these facilities to report a calibrated irradiation setup for both conventional and FLASH studies, as well as a higher extracted beam energy and a larger and more uniform extracted field. We are also currently expanding the external beam therapy capabilities of the BMC to include future proton minibeam (SFRT) studies.

\subsection{Biological results}
As expected, the clonogenic survival assays revealed a higher cytotoxic efficacy of proton irradiation compared to X-rays in both the murine melanoma (B16-F10) cell line and the human keratinocytes (HaCaT). This is supported by the observed \ac{RBE} values of 1.34 for the B16-F10 and 1.21 for the HaCaT cells. The \ac{RBE} values determined in this study are within the range reported in previous studies. For in vitro clonogenic assays, average \ac{RBE} values for protons are generally estimated to be around 1.22 relative to photons \citep{Paganetti2002}. However, most in vitro studies were conducted using V79 hamster fibroblasts and under varied physical conditions, such as proton energies ranging from 65 to 250 MeV \citep{Paganetti2002}, \citep{Paganetti2013}. 
Mara et al \cite{Mara2020} reported an in vitro \ac{RBE} of 1.12 for HaCaT cells at the distal edge of an 80 mm SOBP (\ac{LET} of \SI{4.5}{keV/\micro\meter}). Irradiations were performed with doses of 0.5, 1, 2, 4 and 6 Gy, using a 200 kV X-ray beam as reference. Variations in \ac{RBE} values (\ac{RBE} of 1.21 in our findings compared to the \ac{RBE} of 1.12) are not unexpected and can be attributed to the different experimental conditions, as the \ac{RBE} is known to depend on multiple physical and biological parameters, including dose, dose rate, \ac{LET}, position along the Bragg peak, initial proton energy, the genetic background, and repair capacity of the irradiated cells \citep{Tommasino2015}, \citep{Paganetti2018}.

The survival curves of the HaCaT cell line (Figure \ref{fig:survival_curves}) revealed a relatively shallow shoulder region, with only minor changes in curvature following proton irradiation. This pattern suggests that HaCaT cells possess a limited ability to repair sublethal damage, making them comparatively more radiosensitive. In contrast, the B16-F10 cell line displayed a more pronounced shoulder, indicative of greater sublethal damage repair capacity and, consequently, higher resistance to X-ray radiation.
These findings are consistent with previous reports indicating that rodent cells are generally more resistant to radiation than human cells \citep{Paganetti2002}. 
Despite their greater resistance to X-ray radiation, B16-F10 cells showed increased radiosensitivity after proton irradiation, as shown by the steeper slope of the survival curve, indicative of reduced repair potential and increased damage complexity due to the higher \ac{LET} of protons. Supporting this observation, Belli et al \cite{Belli2000} found that cell lines exhibiting the greatest resistance to photon irradiation showed the greatest increase in radiosensitivity when exposed to higher-\ac{LET} proton radiation.

The $\alpha / \beta$ ratio provides important information on the fractionation sensitivity of tissues and tumours \citep{McMahon2018}. Within the framework of the LQ model, a high $\alpha / \beta$ ratio indicates reduced sensitivity to changes in dose per fraction because cell killing is dominated by the linear component ($\alpha$). Such tissues respond similarly to high and low doses per fraction. In contrast, tissues with a low $\alpha / \beta$ ratio are more sensitive to fraction size, with the quadratic component ($\beta$) playing a greater role \citep{McMahon2018}. 
In the present study, a shift in the $\alpha / \beta$ ratio was observed following proton irradiation (1.09 Gy) compared to X-ray irradiation (3.2 Gy) within the B16-F10 cell line. The lower $\alpha / \beta$ ratio for protons can be attributed to the increased contribution of the $\beta$ component to cell killing (displayed in Table \ref{tab:lq_excel_exact}). 
From a clinical perspective, these findings raise important considerations regarding the optimal fractionation strategy for treating melanoma.
The shift towards a lower $\alpha / \beta$ ratio with proton radiation compared to X-ray therapy suggests that melanoma cells are more sensitive to fraction size with \ac{PT}. 
We hypothesise that \ac{PT} could make the melanoma model more responsive to larger fraction sizes, i.e. higher doses per fraction. This could favour a hypofractionated approach, delivering larger doses per fraction over fewer treatment sessions. This strategy would shorten the overall treatment duration, thereby reducing costs and the burden on patients \citep{Santos2022}. This is particularly relevant given that \ac{PT} is a scarce resource with limited global availability \citep{Santos2022}.
Clinical evidence indicates that cutaneous melanoma is a radioresistant tumour with an $\alpha / \beta$ ratio ranging from 0.6 to 2.5 Gy \citep{Overgaard1986}, \citep{Bentzen1989}, \citep{Strojan2010}. Our results suggest that \ac{PT} could be a more promising and effective option in overcoming the radioresistance of melanoma than conventional radiotherapy. The $\alpha / \beta$ value reported in our study for the mouse melanoma cells falls within the range reported. However, differences between in vitro and in vivo studies, as well as between rodent and human tissues, must be acknowledged. In general, in vitro irradiation assays are useful for characterising cell-intrinsic radiobiological parameters that are important for optimising the therapeutic dose and fractionation scheme \citep{Garate-Soraluze2024}. That said, factors such as the tumour microenvironment and toxicities associated with radiotherapy can only be adequately addressed in vivo.

\section{Conclusion}

This study demonstrates the feasibility of in vitro radiobiological experiments with proton beams at the BMC. RBE values from clonogenic assays agreed with literature, supporting the validity of the irradiation setup and dosimetric procedures. 

The experimental platform developed at the \ac{BMC}, including customized target mounting and real-time dosimetry, provides a robust foundation for pre-clinical radiobiological studies. While further refinement and validation remain warranted, the findings underscore the potential of the \ac{BMC} facility to support advanced proton therapy research, including  future FLASH and \ac{SFRT} studies.

Future work will focus on further improving dosimetric accuracy, improving reproducibility, and expanding the irradiation setup to enable reliable FLASH and \ac{SFRT} dose delivery, as described in Kasanda et al\cite{Kasanda2026-submitted}. These advancements will help establish the \ac{BMC} as a unique and versatile hub for translational proton therapy research.

\acknowledgments
We acknowledge contributions from LHEP engineering and technical staff. This research project was partially funded by the Swiss National Science Foundation (SNSF) (grant: IZURZ2\_224901) and by the BIND Grant programme of the University of Bern. The in-vitro work was part of J. Gruber’s Master thesis \cite{Gruber2026}.

\pagebreak

\section*{Glossary}
\begin{acronym}
  \acro{BMC}{Bern Medical Cyclotron}
  \acro{BTL}{Beam Transfer Line}
  \acro{DBMR}{Department of Biomedical Research}
  \acro{LHEP}{Laboratory for High Energy Physics}
  \acro{LET}{Linear Energy Transfer}
  \acro{LQ}{Linear-Quadratic}
  \acro{PT}{Proton Therapy}
  \acro{RBE}{Relative Biological Effectiveness}
  \acro{RT}{Radiation Therapy}
  \acro{SFRT}{Spatially Fractionated Radiotherapy}
\end{acronym}

\bibliographystyle{JHEP}
\bibliography{mappBib-2}

\providecommand{\href}[2]{#2}\begingroup\raggedright\begin{thebibliography}{10}

\bibitem{Matuszak2022}
N.~Matuszak, W.M.~Suchorska, P.~Milecki, M.~Kruszyna-Mochalska, A.~Misiarz, J.~Pracz et~al., \emph{Flash radiotherapy: an emerging approach in radiation therapy}, \href{https://doi.org/10.5603/rpor.a2022.0038}{\emph{Reports of Practical Oncology and Radiotherapy} {\bfseries 27} (2022) 343–351}.

\bibitem{Mascia2023}
A.E.~Mascia, E.~Anthony, E.C.~Daugherty, Y.~Zhang, E.~Lee, Z.~Xiao et~al., \emph{Proton flash radiotherapy for the treatment of symptomatic bone metastases: The fast-01 nonrandomized trial}, \href{https://doi.org/10.1001/jamaoncol.2022.5843}{\emph{JAMA Oncology} {\bfseries 9} (2023) 62}.

\bibitem{Yan2020}
W.~Yan, M.K.~Khan, X.~Wu, C.B.~Simone, J.~Fan, E.~Gressen et~al., \emph{Spatially fractionated radiation therapy: History, present and the future}, \href{https://doi.org/10.1016/j.ctro.2019.10.004}{\emph{Clinical and Translational Radiation Oncology} {\bfseries 20} (2020) 30–38}.

\bibitem{Mali2024}
S.B.~Mali, \emph{Mini review of spatially fractionated radiation therapy for cancer management}, \href{https://doi.org/10.1016/j.oor.2024.100175}{\emph{Oral Oncology Reports} {\bfseries 9} (2024) 100175}.

\bibitem{Kasanda2026-submitted}
E.~Kasanda, L.~Eggiman, T.~Stammbach, P.~Casolaro, G.~Dellepiane, A.~Gottstein et~al., \emph{Development of a proton therapy research beamline with flash and minibeam capabilities at the 18 mev bern medical cyclotron},  2026.
\newblock 10.48550/ARXIV.2605.05441.

\bibitem{Braccini2013}
S.~Braccini, \emph{The new bern pet cyclotron, its research beam line, and the development of an innovative beam monitor detector},  in \emph{AIP Conference Proceedings}, AIP, 2013, \href{https://doi.org/10.1063/1.4802308}{DOI}.

\bibitem{Auger2015}
M.~Auger, S.~Braccini, A.~Ereditato, K.P.~Nesteruk and P.~Scampoli, \emph{Low current performance of the bern medical cyclotron down to the pa range}, \href{https://doi.org/10.1088/0957-0233/26/9/094006}{\emph{Measurement Science and Technology} {\bfseries 26} (2015) 094006}.

\bibitem{Potkins2017}
D.E.~Potkins, S.~Braccini, K.P.~Nesteruk, T.S.~Carzaniga, A.~Vedda, N.~Chiodini et~al., \emph{A low-cost beam profiler based on cerium-doped silica fibers}, \href{https://doi.org/10.1016/j.phpro.2017.09.061}{\emph{Physics Procedia} {\bfseries 90} (2017) 215–222}.

\bibitem{Gottstein2025}
A.~Gottstein, L.~Mercolli, E.~Kasanda, I.~Mateu, L.~Eggimann, E.~Zyaee et~al., \emph{Beam energy measurement using a bayesian approach with the stacked foil method},  2025.
\newblock 10.48550/ARXIV.2510.14440.

\bibitem{Tarasov2008}
O.~Tarasov and D.~Bazin, \emph{Lise++: Radioactive beam production with in-flight separators}, \href{https://doi.org/10.1016/j.nimb.2008.05.110}{\emph{Nuclear Instruments and Methods in Physics Research Section B: Beam Interactions with Materials and Atoms} {\bfseries 266} (2008) 4657–4664}.

\bibitem{PTW}
{PTW Freiburg GmbH}, \emph{Monitor ionization chambers 34014, 786},  2025.

\bibitem{Ashland2023}
{Ashland Advanced Materials}, \emph{EBT-3 FILM SPECIFICATION AND USER GUIDE}.
\newblock Wilmington, 8145 Blazer Dr, United States, 2023.

\bibitem{SanchezParcerisa2021}
D.~Sanchez-Parcerisa, I.~Sanz-García, P.~Ibáñez, S.~España, A.~Espinosa, C.~Gutiérrez-Neira et~al., \emph{Radiochromic film dosimetry for protons up to 10 mev with ebt2, ebt3 and unlaminated ebt3 films}, \href{https://doi.org/10.1088/1361-6560/abfc8d}{\emph{Physics in Medicine \& Biology} {\bfseries 66} (2021) 115006}.

\bibitem{Gruber2026}
J.~Gruber, \emph{Establishment of the first biological experiment using proton particles at the Bern Medical Cyclotron and it's Beam Transfer Line}, master’s thesis, University of Bern, 2026.
\newblock 10.5281/ZENODO.21376885.

\bibitem{Boukamp1988}
P.~Boukamp, R.T.~Petrussevska, D.~Breitkreutz, J.~Hornung, A.~Markham and N.E.~Fusenig, \emph{Normal keratinization in a spontaneously immortalized aneuploid human keratinocyte cell line.}, \href{https://doi.org/10.1083/jcb.106.3.761}{\emph{The Journal of cell biology} {\bfseries 106} (1988) 761–771}.

\bibitem{Nakamura2002}
K.~Nakamura, N.~Yoshikawa, Y.~Yamaguchi, S.~Kagota, K.~Shinozuka and M.~Kunitomo, \emph{Characterization of mouse melanoma cell lines by their mortal malignancy using an experimental metastatic model}, \href{https://doi.org/10.1016/s0024-3205(01)01454-0}{\emph{Life Sciences} {\bfseries 70} (2002) 791–798}.

\bibitem{Ohira1994}
T.~Ohira, Y.~Ohe, Y.~Heike, E.R.~Podack, K.J.~Olsen, K.~Nishio et~al., \emph{In vitro and in vivo growth of b16f10 melanoma cells transfected with interleukin-4 cdna and gene therapy with the transfectant}, \href{https://doi.org/10.1007/bf01245372}{\emph{Journal of Cancer Research and Clinical Oncology} {\bfseries 120} (1994) 631–635}.

\bibitem{Yerlikaya2008}
A.~Yerlikaya and N.~Erin, \emph{Differential sensitivity of breast cancer and melanoma cells to proteasome inhibitor velcade}, \href{https://doi.org/10.3892/ijmm\_00000090}{\emph{International Journal of Molecular Medicine} {\bfseries 22} (2008) 817}.

\bibitem{Fabbrizi2025}
M.R.~Fabbrizi, J.R.~Hughes, L.D.~Punshon, L.~Hawkins, V.~Sorokin, A.~Ormrod et~al., \emph{Radiobiological characterisation of a 28 mev proton beam delivered by the mc-40 cyclotron}, \href{https://doi.org/10.1038/s41420-025-02635-1}{\emph{Cell Death Discovery} {\bfseries 11} (2025) }.

\bibitem{Evin2024}
M.~Evin, C.~Koumeir, A.~Bongrand, G.~Delpon, F.~Haddad, Q.~Mouchard et~al., \emph{Methodology for small animals targeted irradiations at conventional and ultra-high dose rates 65 mev proton beam}, \href{https://doi.org/10.1016/j.ejmp.2024.103332}{\emph{Physica Medica} {\bfseries 120} (2024) 103332}.

\bibitem{Ralite2026}
F.~Ralite, M.~Evin, C.~Koumeir, A.~Guertin, F.~Haddad, Q.~Mouchard et~al., \emph{Proton beam monitoring through water scintillation in radiobiology experiments}, \href{https://doi.org/10.1016/j.radphyschem.2025.113353}{\emph{Radiation Physics and Chemistry} {\bfseries 239} (2026) 113353}.

\bibitem{jung2012}
U.~Jung, H.S.~Eom, K.~Jeong, H.-R.~Park and S.-K.~Jo, \emph{In vitro and in vivo {Biological} {Responses} of {Proton} {Irradiation} from {MC}-50 {Cyclotron}}, \href{https://doi.org/10.23042/radin.2012.6.3.223}{\emph{Journal of Radiation Industry} {\bfseries 6} (2012) 223}.

\bibitem{BarattoRoldn2020}
A.~Baratto-Roldán, M.d.C.~Jiménez-Ramos, S.~Jimeno, P.~Huertas, J.~García-López, M.I.~Gallardo et~al., \emph{Preparation of a radiobiology beam line at the 18 mev proton cyclotron facility at cna}, \href{https://doi.org/10.1016/j.ejmp.2020.04.022}{\emph{Physica Medica} {\bfseries 74} (2020) 19–29}.

\bibitem{Teixeira2026}
A.R.C.~Teixeira, S.J.~Do~Carmo, S.~Silva, C.I.~Pinto, P.~Santos, P.~Crespo et~al., \emph{Characterization of a low-energy cyclotron-based proton beam for preclinical radiobiological studies}, \href{https://doi.org/10.1016/j.ijpt.2026.101318}{\emph{International Journal of Particle Therapy} {\bfseries 20} (2026) 101318}.

\bibitem{Paganetti2002}
H.~Paganetti, A.~Niemierko, M.~Ancukiewicz, L.E.~Gerweck, M.~Goitein, J.S.~Loeffler et~al., \emph{Relative biological effectiveness (rbe) values for proton beam therapy}, \href{https://doi.org/10.1016/s0360-3016(02)02754-2}{\emph{International Journal of Radiation Oncology*Biology*Physics} {\bfseries 53} (2002) 407–421}.

\bibitem{Paganetti2013}
H.~Paganetti and P.~van Luijk, \emph{Biological considerations when comparing proton therapy with photon therapy}, \href{https://doi.org/10.1016/j.semradonc.2012.11.002}{\emph{Seminars in Radiation Oncology} {\bfseries 23} (2013) 77–87}.

\bibitem{Mara2020}
E.~Mara, M.~Clausen, S.~Khachonkham, S.~Deycmar, C.~Pessy, W.~D\"{o}rr et~al., \emph{Investigating the impact of alpha/beta and letd on relative biological effectiveness in scanned proton beams: An in vitro study based on human cell lines}, \href{https://doi.org/10.1002/mp.14212}{\emph{Medical Physics} {\bfseries 47} (2020) 3691–3702}.

\bibitem{Tommasino2015}
F.~Tommasino and M.~Durante, \emph{Proton radiobiology}, \href{https://doi.org/10.3390/cancers7010353}{\emph{Cancers} {\bfseries 7} (2015) 353–381}.

\bibitem{Paganetti2018}
H.~Paganetti, \emph{Proton relative biological effectiveness – uncertainties and opportunities}, \href{https://doi.org/10.14338/ijpt-18-00011.1}{\emph{International Journal of Particle Therapy} {\bfseries 5} (2018) 2–14}.

\bibitem{Belli2000}
M.~Belli, D.~Bettega, P.~Calzolari, F.~Cera, R.~Cherubini, M.D.~Vecchia et~al., \emph{Inactivation of human normal and tumour cells irradiated with low energy protons}, \href{https://doi.org/10.1080/09553000050028995}{\emph{International Journal of Radiation Biology} {\bfseries 76} (2000) 831}.

\bibitem{McMahon2018}
S.J.~McMahon, \emph{The linear quadratic model: usage, interpretation and challenges}, \href{https://doi.org/10.1088/1361-6560/aaf26a}{\emph{Physics in Medicine \& Biology} {\bfseries 64} (2018) 01TR01}.

\bibitem{Santos2022}
A.~Santos, S.~Penfold, P.~Gorayski and H.~Le, \emph{The role of hypofractionation in proton therapy}, \href{https://doi.org/10.3390/cancers14092271}{\emph{Cancers} {\bfseries 14} (2022) 2271}.

\bibitem{Overgaard1986}
J.~Overgaard, M.~Overgaard, P.~Vejby~Hansen and H.~von~der Maase, \emph{Some factors of importance in the radiation treatment of malignant melanoma}, \href{https://doi.org/10.1016/s0167-8140(86)80048-2}{\emph{Radiotherapy and Oncology} {\bfseries 5} (1986) 183–192}.

\bibitem{Bentzen1989}
S.~Bentzen, J.~Overgaard, H.~Thames, M.~Overgaard, P.~Hansen, H.~von~der Maase et~al., \emph{Clinical radiobiology of malignant melanoma}, \href{https://doi.org/10.1016/0167-8140(89)90017-0}{\emph{Radiotherapy and Oncology} {\bfseries 16} (1989) 169–182}.

\bibitem{Strojan2010}
P.~Strojan, \emph{Role of radiotherapy in melanoma management}, \href{https://doi.org/10.2478/v10019-010-0008-x}{\emph{Radiology and Oncology} {\bfseries 44} (2010) 1–12}.

\bibitem{Garate-Soraluze2024}
E.~Garate-Soraluze, J.~Marco-Sanz, I.~Serrano-Mendioroz, L.~Marrodán, L.~Fernandez-Rubio, S.~Labiano et~al., \emph{Chapter seven - radiotherapy protocols for mouse cancer model},  in \emph{Animal Models of Disease - Part A}, J.M.~{Bravo-San Pedro}, F.~Aranda, A.~Buqué and L.~Galluzzi, eds., vol.~185 of \emph{Methods in Cell Biology}, pp.~99--113, Academic Press (2024), \href{https://doi.org/https://doi.org/10.1016/bs.mcb.2024.02.007}{DOI}.

\end{thebibliography}\endgroup






\end{document}